\documentclass[letterpaper]{jpconf}
\usepackage{graphicx}
\begin{document}
\title{The role of collisions and strong coupling in ultracold plasmas}

\author{ J  Castro, P  McQuillen, H  Gao  and T  C  Killian}

\address{Rice University, Department of Physics and Astronomy, 6100 Main St., Houston, Texas, USA 77005}

\ead{killian@rice.edu}

\begin{abstract}
Ultracold plasmas are formed by photo-exciting clouds of cold atoms and
molecules near the ionization threshold. They explore a new region of plasma
physics and display effects of strong coupling, which is characterized by a
ratio of Coulomb energy to kinetic energy that is greater than unity.
Collisions of many types play a role in the creation, equilibration, and
expansion of these systems.\end{abstract}

\section{Introduction}

Ultracold neutral plasmas are formed by photo-exciting cold atoms and molecules
near the ionization threshold, yielding electron temperatures from 1-1000K and
 ion temperatures around 1\,K \cite{kil07,kpp07}. At typical densities of
$\sim 10^{10}$\,cm$^{-3}$, the average Coulomb interaction energy between
neighboring particles can be on the order of or exceed the thermal energy,
which makes the plasma strongly coupled \cite{ich82}. Strong coupling is
relatively difficult to achieve experimentally, but it is important in many
areas of physics. In ultracold plasmas it leads to spatial correlations,
modification of thermalization timescales, and surprising equilibration
dynamics. Ultracold plasmas also display many classic plasma phenomena such as
various collective modes, Debye screening, and ambipolar diffusion.

\begin{figure}
 \mbox{\small
 \begin{minipage}[b]{6.15in}
  \begin{minipage}[t]{2.15in}
        \mbox{}\\ \unitlength0.75cm
         \includegraphics[width=1.75in,clip=true,angle=270,trim=00 00 300 00]{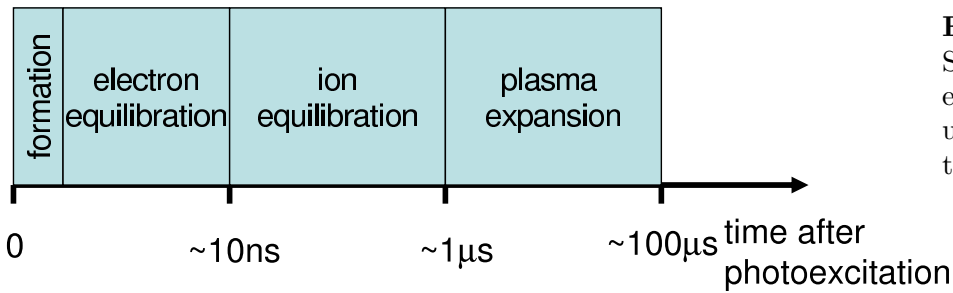}\hspace{-0pc}\\
  \end{minipage} \hfill
    \raisebox{-.25in}{\parbox[t]{1in}{\makebox[0cm]{}{ \caption{Stages in the evolution of an ultracold
neutral plasma.}\label{Figure: timeline} }}}
 \end{minipage}}
\end{figure}

Collisions of many types play prominent roles in the main stages of plasma
evolution: formation, electron equilibration, ion equilibration, and plasma
expansion (Fig.\ \ref{Figure: timeline}). When starting with a gas of highly
excited Rydberg atoms, atom-atom and electron-atom collisions lead to
ionization and plasma formation. Electron collisions, three-body recombination,
and electron-Rydberg collisions dominate dynamics during electron
equilibration. During the ion equilibration stage, ions heat due to
interactions with neighboring ions, and collisional processes reflect strong
coupling of the ions. During expansion,  electrons and ions adiabatically cool,
which eventually increases the three-body recombination rate. Formation of the
plasma from molecules in a beam introduces predissociation, dissociative
recombination, and collisions with neutral atoms forming the high pressure
backing gas.


\section{Strong Coupling}
In strongly coupled plasmas, the Coulomb interaction energy exceeds the thermal
energy, and this is parameterized by the Coulomb coupling constant
\begin{equation}\label{eqcoulomncouplingconstant}
    \Gamma=\frac{e^{2}}{4 \pi \varepsilon_0 a k_B T},
\end{equation}
where $T$ is the temperature and $a$ is the Wigner-Seitz radius,
$a=\left[3/(4\pi n)\right]^{1/3}$ for density $n$.  Ions in ultracold neutral
plasmas equilibrate with $2< \Gamma_i < 5$ \cite{scg04}. It is possible to set
initial conditions that imply large electron $\Gamma_e$, but rapid heating
processes involving many different collisional effects clamp the equilibrium
$\Gamma_e \leq 0.2$ \cite{rha03,kon02,mck02,gls07}.

Strong coupling is of interest in many areas of physics. In systems with
$\Gamma>1$, spatial correlations develop between particles, concepts such as
Debye screening and hydrodynamics must be reexamined, and new phenomena appear,
such as kinetic energy oscillations during equilibration \cite{csl04}.
Strongly-coupled plasmas appear in extreme environments, such as dense
astrophysical systems \cite{vho91}, matter irradiated with intense-laser fields
\cite{nmg98, sht00}, dusty plasmas of highly charged macroscopic particles
\cite{mtk99}, or non-neutral trapped ion plasmas \cite{mbh99} that are
 laser-cooled until they freeze into a Wigner crystal. Ultracold neutral
 plasmas offer unique opportunities for studying strong coupling, especially
 because the method of forming the plasma allows equilibration processes to be
 studied in great detail.

\section{Creation of an Ultracold Neutral Plasma}

\subsection{Direct Photoionization of Laser-Cooled Atoms}
The most common method for creating an ultracold neutral plasma is to start
with laser-cooled and trapped neutral atoms in a magneto-optical trap
(MOT)\cite{mvs99}. Most experiments use alkali metal atoms, alkaline-earth
metal atoms, or metastable noble gas atoms because their principal transitions
are at convenient laser wavelengths. Depending upon the element chosen, up to
$10^{10}$ atoms can be trapped, and the density can be as high as $10^{12}
\,\,{\rm cm}^{-3}$, but most experiments are conducted with about 100 times
lower values. The temperature is in the microkelvin to millikelvin range. The
density distribution is typically a spherical Gaussian, with a characteristic
radius of about 1 mm, which also determines the density profile of the plasma.


\begin{figure}[h]
\includegraphics[width=2.5in,clip=true,trim=120 0 00 000,angle=270]{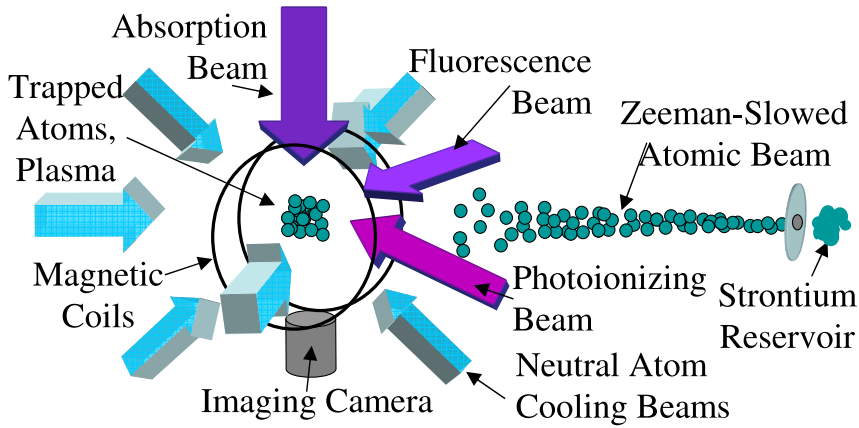}\hspace{-2pc}%
\caption{\label{Figure: fig1apparatus} Typical experimental schematic for
creating an ultracold plasma from
    laser cooled atoms in a MOT \cite{scg04}.
   The MOT for neutral atoms consists of a pair of anti-Helmholtz magnetic
  coils and 6  laser-cooling beams. Atoms from a Zeeman-slowed
  atomic beam enter the MOT region, are trapped, and then ionized by the photoionizing
  laser. Optical imaging diagnostics, such as absorption and fluorescence imaging, can
  be used for alkaline-earth metal plasmas.
  For charged particle detection diagnostics, electron or ion multipliers
and electric field-generating wires or meshes are added to the vacuum system to
detect charges that escape from  the plasma.}
\end{figure}

The plasma is created through single or multi-photon processes usually
involving  a $10$~ns pulsed dye laser whose wavelength is tuned just above the
ionization continuum, and the ionization fraction can approach $75$\%. Because
of the small electron-ion mass ratio, the electrons have an initial kinetic
energy approximately equal to the difference between the photon energy and the
ionization potential, typically between 1 and $1000$\,K. The initial kinetic
energy for the ions is close to the kinetic energy of neutral atoms in the MOT.

\subsection{Plasma Creation from Rydberg Atoms}

If the ionizing laser wavelength is tuned just below the ionization threshold,
a dense gas of highly excited Rydberg atoms is created. It was noted very early
in the study of these systems that in a low electric field environment, in
which free electrons are not quickly removed from the system, electron-atom
collisions can drive a spontaneous ionization cascade \cite{rbs98,rtn00}.
Approximately two-thirds of the atoms are ionized, and the remaining fraction
are driven to more deeply bound states, satisfying conservation of energy.

This formation process is quite complex. Initial electrons are produced by
blackbody ionization \cite{rtn00} or ionizing collisions of initially
stationary atoms that are driven by resonant dipole-dipole attractive
interactions \cite{ltg05}. Electrons also efficiently drive angular
momentum-changing collisions and rapidly mix the $\ell$ values of the Rydberg
atoms \cite{dfw01}.

It is interesting to note that independent of the creation technique, during
the expansion of ultracold plasmas, a significant fraction ($\geq 10$\%) of the
charges recombine to form Rydberg atoms \cite{klk01}, presumable through
three-body recombination \cite{rha02,gls07}.

\subsection{Plasma Creation from Molecular Beams}
A recent advance in the field, which greatly expands the types of plasmas that
can be formed and the collisional processes present is the creation of an
ultracold plasma in a supersonic molecular beam \cite{mrk08}. The effective
translational temperature in the moving frame of reference of atoms or
molecules seeded in the expansion of a high pressure ($\geq$ 1 atm), typically
inert, backing gas can be in the range of 1 K. Rotational and vibrational
temperatures can also be reduced. Such  beams have been valuable tools in
chemistry and physics for many years (e.g.\cite{glh71}) because of the utility
of the cold samples for experiments and general applicability to any atomic and
molecular species.

A supersonic beam is an attractive starting point for forming an ultracold
plasma because the density of target species can be high ($\geq
10^{12}$\,cm$^{-3}$), photoexcitation of a well defined volume of gas is easily
accomplished with pulsed lasers, and it gives easy access to cold molecules,
which are currently challenging to produce in any other way at high density.
Morrison \textit{et al.} \cite{mrk08} excited a dense gas of NO Rydberg
molecules in this way and observed the spontaneous evolution into a plasma,
exactly as observed in \cite{rbs98,rtn00}. A time-of-flight charged particle
diagnostic was used to demonstrate that a stable, quasi-neutral plasma was
formed. The rate of plasma expansion was used to infer that the plasma was
ultracold, as expected, with an electron temperature as a low as 7\,K
\cite{mrg09}. This experiment raises many exciting possibilities, such as
further cooling of ions in the plasma through collisions with the background
gas and the introduction of new collisional phenomena, such as predissociation
and dissociative recombination, which may affect plasma dynamics.

\section{Diagnostics}
A full description of UNP diagnostic techniques can be found in \cite{kpp07}.
Charged particle detection of ions or electrons is the most common technique
because of its simplicity and applicability to any type of plasma. Information
can be extracted on the rate of escape of electrons from the plasma
\cite{kkb99}, the spatial distribution of ions \cite{mrk08}, and the response
of the plasma to external perturbation \cite{fzr07}, such as a radio-frequency
field. In conjunction with pulsed-field ionization, charged particle
diagnostics can be used to measure Rydberg atom populations in the plasma
\cite{dfw01,ltg05}. Optical imaging techniques, which use lasers resonant with
a principal transition in the ions, have proven very powerful because they can
measure the ion kinetic energy and temperature \cite{csl04,cdd05}. But they
have only been used with plasmas formed from alkaline-earth metal atoms, which
have ions with principal transitions in the visible. Fluorescence imaging has
recently been used to separate the large ion kinetic energy due to plasma
expansion from thermal kinetic energy \cite{kgc08}.

\section{Electron Equilibration}

When the plasma is formed, both electrons and ions are far from equilibrium,
and most ultracold neutral plasma studies have focused on the establishment of
local and global thermal equilibrium. This topic takes on particular interest
because the plasma is in or near the strongly coupled regime. For a thorough
review, see \cite{kpp07}.

\subsection{Disorder-Induced Heating}

As the name suggests, disorder-induced heating (DIH) stems from the fact that
immediately after plasma creation, charged particles are spatially
uncorrelated, which yields  greater potential energy than the equilibrium
state. On a timescale of the inverse electron plasma frequency,
 $\omega_{pe}^{-1}=\sqrt{m_e
\varepsilon_0/n_e e^2}$, where $m_e$ is the electron mass and $e$ is the
electron charge, which is on the order of a few nanoseconds,
 short range correlations develop between electrons and ions and electrons and
 electrons. This
 decreases the potential energy and  increases the kinetic energy by about
 $e^2/(4\pi \varepsilon_0 a)$ per particle,
 which heats electrons typically by a few degrees kelvin \cite{gls07}.
 Kuzmin et al. \cite{kon02prl} showed that
 correlations develop until the Coulomb coupling parameter approaches $\Gamma_e\sim1$.

\subsection{Coulomb Collisions}

During DIH, the electron velocity distribution is also thermalizing due to
Coulomb collisions. The timescale for this process is given by the
Spitzer-Landau formula \cite{Spitzer},
\begin{eqnarray}\label{spitzerlandau}
\tau_{ee}\sim\frac{2\pi\varepsilon_0^2m_e^{1/2}(3k_BT_e)^{3/2}}{n_ee^4ln(\Lambda_e)}, 
\end{eqnarray}
where $ln(\Lambda_e)=ln(4\pi\varepsilon_03k_BT_e\lambda_{D,e}/e^2)$ is the so-called
Coulomb logarithm for the electrons with $\lambda_{D,e}=\sqrt{\varepsilon_0k_BT_e/(n_ee^2)}$ as the
electron Debye screening length.
 For $T_e = 40$\,K and $n_e = 10^{15}$\,m$^{-3}$,
$\tau_{ee}\sim10$\,ns. After thermalization, they can be described with a
Maxwell-Boltzmann distribution and a well-defined local temperature
\cite{rha03,ppr04archive}.

The mean-free-path for electrons is longer than the plasma size, so the
electrons are in good thermal contact with each other. Nonetheless there can be
small variations in electron temperature due to the spatial dependence of the
potential seen by electrons and the fact that electrons escape from the edge of
the plasma \cite{rha03}. Many papers have studied various aspects of the
electron dynamics \cite{rha02,rha03,kon02,mck02,gls07} such as ambipolar
diffusion and trapping of electrons in the potential formed by the ions
\cite{kkb99}.

\subsection{Three-Body Recombination}

Three-body recombination(TBR) \cite{mke69} refers to the process in which an
ion and electron recombine to form highly excited Rydberg atoms and the energy
released in this process is taken up by a second electron to conserve energy
and momentum.
The rate of TBR, per ion \cite{mke69,kon02}, is given as
\begin{eqnarray}\label{eqTBR}
R
=3.9\times10^{-21}\mathrm{s}^{-1}
\left[n_e(\mathrm{m}^{-3})\right]^2\left[T_e(\mathrm{K})\right]^{-9/2}.
\end{eqnarray}
As the TBR rate varies with electron temperature as $T_e^{-9/2}$, this is the dominant recombination mechanism in UNPs \cite{klk01}.  Because TBR acts as a heating mechanism for the
electrons and the rate increases so rapidly with decreasing temperature,
TBR acts as a natural feedback mechanism, which has been shown experimentally
and theoretically to keep the $\Gamma_e \leq 0.2$
\cite{rha03,kon02,mck02,gls07}.

%
There has been much debate about the validity of the TBR expression, Eq.
\ref{eqTBR}, for ultracold plasmas in which the electron temperatures are very
low, e.g. \cite{hah97,mck02,kon02prl,rha03,fzr07}. A recent Monte Carlo
calculation appropriate to conditions in an UNP \cite{pvs08} found slight
deviations from the classic expressions \cite{mke69} at very low collision
energies, which decreases the overall rate by 30\% but does not qualitatively
change the behavior. This discussion is complicated by the challenge of
describing rates in a system that is, strictly speaking, not in equilibrium.
Also, electrons and ions are constantly forming weakly bound states that are
quickly disrupted by plasma microfields, so even the definition of a Rydberg
atom can be called into question. But a convergence seems to be emerging around
theory \cite{pvs08} and experimental results measuring the rate at which
Rydberg states are repopulated after being emptied by an RF pulse \cite{fzr07}.


\subsection{Rydberg-electron collisions}
A collision between a Rydberg atom and an electron in the plasma tends to
ionize the Rydberg atom if it is bound by less than about $4k_BT$ and drive it
to lower energies if it is bound by more than this. This leads to a ``kinetic
bottleneck" in the population \cite{pvs08}, but it also leads to significant
further heating of the plasma electrons \cite{rha03,gls07} since a tremendous
amount of energy can be released during de-excitation. It has also been shown
that electron-Rydberg collisions in UNPs are extremely efficient at randomizing
the distribution of occupied electron orbital angular momentum states
\cite{dfw01}.

\section{Ion Equilibration}

The ions in strontium UNPs are created with kinetic energies reflecting the
millikelvin or microkelvin temperature of atoms in the MOT. This would suggest
very large $\Gamma_i$ and very strong spatial correlations. But immediately
after plasma formation, the ions are far from thermal equilibrium, and the
equilibration process drastically changes the kinetic energy distribution and
displays phenomena that reflect strong coupling.


\subsection{Disorder-Induced Heating}
DIH affects the ions just as it does the electrons. It raises the ion
temperature to about 1\,K (Fig.\ \ref{figkeo}) and acts as a feedback mechanism
to always produce $2<\Gamma_i<5$ \cite{scg04,csl04}. The timescale, given by
the inverse ion plasma oscillation frequency
$\omega_{pi}^{-1}=\sqrt{m_i\varepsilon_0/n_{i}e^2}\sim 1$\,$\mu$s, where,
$n_{i}$ is the ion-density and $m_i$ is the mass of the ion, is much slower
than for electrons.  Debye screening by the electrons modifies the equilibrium
ion temperature, and this effect is well described by modelling the ion-ion
interaction as a Yukawa interaction \cite{mur01} even when there are only a few
electrons per Debye sphere \cite{csl04}.

\subsection{Kinetic Energy Oscillations}
\begin{figure}
 \mbox{\small
 \begin{minipage}[b]{6.0in}
  \begin{minipage}[t]{1in}
        \mbox{}\\ \unitlength0.75cm
           \includegraphics[width=3.5in,clip=true,trim=0 0 50 0]{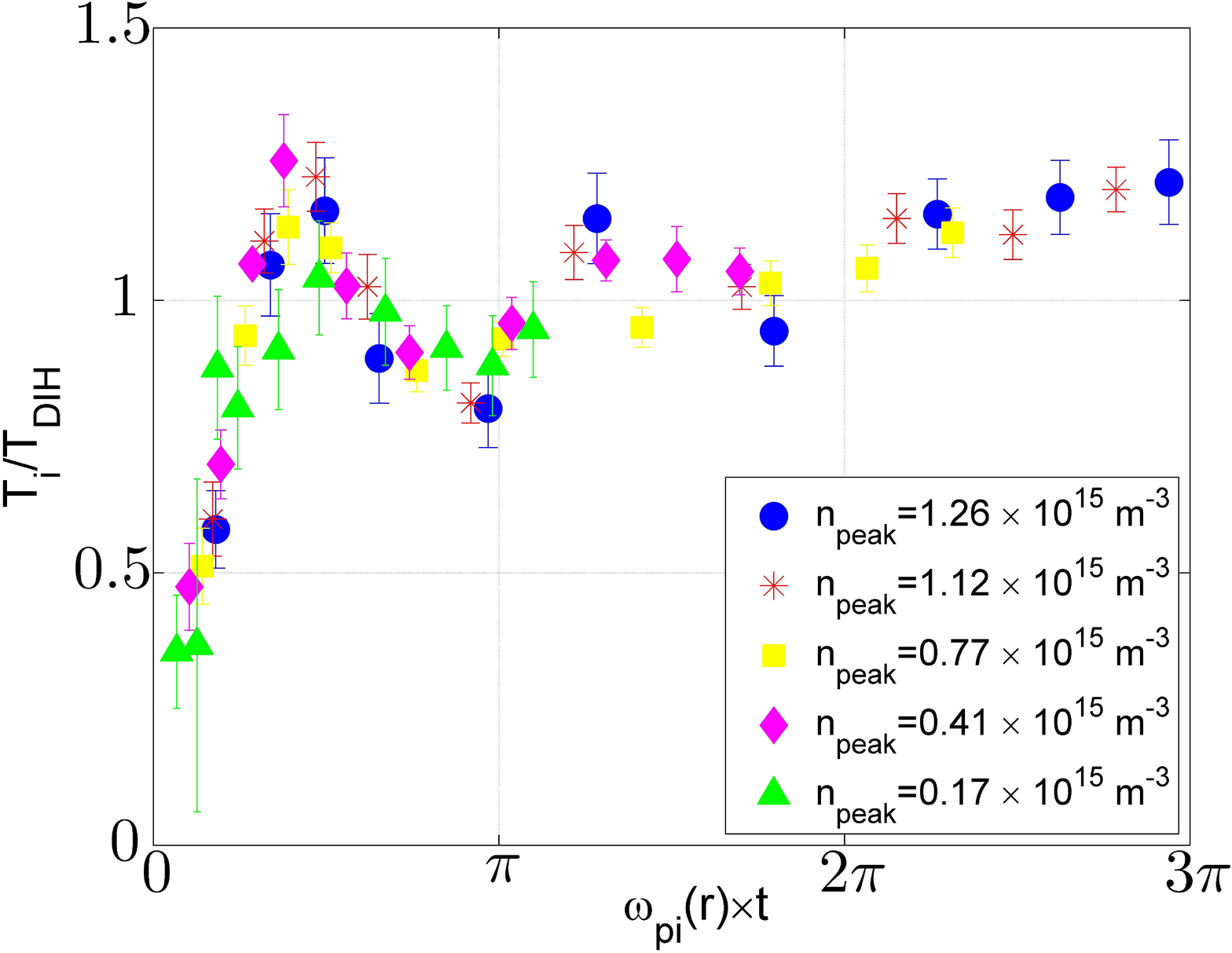}\\
  \end{minipage} \hfill
  \raisebox{-.15in}{\parbox[t]{2.25in}{\makebox[0cm]{}{\caption{ Disorder-induced heating and kinetic energy oscillations in a UNP.
As spatial correlations develop, ion potential energy is converted to kinetic
energy to a value given by $T_{DIH}\sim$1\,K \cite{mur01}.  The kinetic energy
will oscillate about its equilibrium value with the plasma frequency,
$\omega_{pi} \propto n^{1/2}$.  The figure shows how disorder-induced heating
and the oscillation frequency scale for UNPs with different
densities.}\label{figkeo} }}}
 \end{minipage}}
\end{figure}

During equilibration, strong coupling gives rise to oscillations in the kinetic
energy at $2\omega_{pi}$ reflecting each ion's oscillation at frequency
$\omega_{pi}$ in the potential well formed by the cage of nearest neighbors.
Approximately one coherent oscillation is observable, before collisional
processes and nonuniformity of the potential wells seen by each ion, damp the
oscillation. DIH can be viewed as the beginning of this oscillation. This
phenomenon was first observed experimentally in \cite{csl04} and modeled more
extensively in \cite{lcg06,mur06PRL}, and similar dynamics should occur in
plasmas produced by intense radiation of solid and foil targets
\cite{mur06PRL}.  Although there is no long-range spatial coherence in the
oscillation, it is instructive to think of this motion as an ion plasma
oscillation, which is the short wavelength limit of the ion electrostatic wave. Figure \ref{figkeo} shows kinetic energy oscillations for UNPs with different densities and its scaling with density.


\subsection{Coulomb Collisions}

Traditionally, the equilibration time scale for  the ion velocity distribution is calculated with the Spitzer-Landau formula \cite{Spitzer},
\begin{eqnarray}
\tau_{ii}\sim\frac{2\pi\varepsilon_0^2m_i^{1/2}(3k_BT_i)^{3/2}}{n_ie^4ln(\Lambda_i)},
\end{eqnarray}
where, for the ions and for $T_i\ll T_e$, the Coulomb logarithm is now
evaluated as $ln(\Lambda_i)=ln(4\pi\varepsilon_03k_BT_i\lambda_{D,i}/e^2)$ with
the ion Debye screening length of
$\lambda_{D,i}=\sqrt{\varepsilon_0k_BT_i/(n_ie^2)}$. In terms of the coupling
parameter $\Gamma$, the logarithm may be written as
$ln(\Lambda_i)=ln(\sqrt{3}/\Gamma_i^{3/2})$.  This expression diverges when the
temperature and density correspond to a strongly coupled plasma, $\Gamma_i>1$.



To understand this the origin of this divergence, note that the Spitzer-Landau
time is calculated from the energy exchange time in a collision between charged
partilces \cite{Spitzer}. This calculation involves the velocity diffusion
coefficients  found by integrating over all possible impact parameters, $\rho$.
This integration diverges due to contributions of large impact parameters.
Canonically, this divergence is removed by introducing
an upper limit cut-off at $\rho_{max}= \lambda_{D,i}$, the ion Debye screening
length. This is a physically sensible cutoff in weakly coupled plasmas in which
the there are many particles within a sphere of radius $\lambda_{D}$.
But in strongly coupled systems, as the
Debye screening length becomes similar to the interparticle spacing,
it is no longer suitable as the upper limit cut-off, and the Spitzer formula
becomes invalid.

Many alternative cutoffs and expressions for $\tau_{ii}$ have been proposed to
remove the divergence of the Coulomb logarithm in strongly coupled systems
\cite{gms02}. This is an important problem since Coulomb collisions and
associated relaxation times are crucial to inertial confinement fusion
experiments \cite{ggm08}. UNPs can thus provide a good tool to study this
problem.

\subsection{Collisions with Background Neutral Atoms}
Collisions between ions and neutral atoms and molecules at ultracold
temperatures \cite{rda00} have attracted increased attention recently, in part
because of experiments creating overlapping trapped atoms and ions
\cite{gco09}. Ultracold plasmas formed by photoexciting molecules seeded in a
supersonic beam \cite{mrk08} provide a new opportunity to study such phenomena,
and the collisions may serve a valuable function by removing disorder-induced
heat from the ions to produce much stronger coupling. Large elastic collision
cross-sections, $\sigma\sim 10^{-10}$\,cm$^{2}$ \cite{rda00}, at collision
energies near 1\,K  lead to collision times below 1\,ms for a background gas
density $\sim 10^{12}$\,cm$^{-3}$.

\section{Conclusion}
With the introduction of new diagnostics, such as fluorescence imaging, and new
techniques, such as formation of plasmas in supersonic beams, the range of
phenomena accessible for study in UNPs continues to expand. Collisions of many
types are central to the dynamics of these system, and strong coupling adds
interesting new features to familiar processes such as equilibration.

\ack This work is supported by the U.S. National Science Foundation-DOE
Partnership and the David and Lucille Packard Foundation. J.C. thanks MICIT and
CONICIT of Costa Rica for financial support.

\providecommand{\newblock}{}


\end{document}